\documentclass{article}

\PassOptionsToPackage{square, numbers}{natbib}

\usepackage[preprint]{neurips_2024}

\usepackage[utf8]{inputenc} 
\usepackage[T1]{fontenc}    
\usepackage{hyperref}       
\usepackage{url}            
\usepackage{booktabs}       
\usepackage{amsfonts}       
\usepackage{nicefrac}       
\usepackage{microtype}      
\usepackage{xcolor}         
\usepackage{bm}
\usepackage{graphicx}
\usepackage{makecell}
\usepackage{multirow,multicol}
\usepackage{threeparttable}
\usepackage{bbding}
\usepackage{booktabs}
\title{Wav2Prompt: End-to-End Speech Prompt Generation and Tuning For LLM in Zero and Few-shot Learning}

\author{%
  Keqi Deng, Guangzhi Sun, Philip C. Woodland\\
  Department of Engineering, University of Cambridge\\
  Trumpington St., Cambridge, UK \\
  \texttt{\{kd502, gs534, pw117\}@cam.ac.uk} \\
}

\begin{document}

\maketitle

\begin{abstract}

Wav2Prompt is proposed which allows straightforward integration between spoken input and a text-based large language model (LLM). Wav2Prompt uses a simple training process with only the same data used to train an automatic speech recognition (ASR) model. After training, Wav2Prompt learns continuous representations from speech and uses them as LLM prompts.  
To avoid task over-fitting issues found in prior work and preserve the emergent abilities of LLMs, Wav2Prompt takes LLM token embeddings as the training targets and utilises a continuous integrate-and-fire mechanism for explicit speech-text alignment.
Therefore, a Wav2Prompt-LLM combination can be applied to zero-shot spoken language tasks such as speech translation (ST), speech understanding (SLU), speech question answering (SQA) and spoken-query-based QA (SQQA). It is shown that for these zero-shot tasks, Wav2Prompt performs similarly to an ASR-LLM cascade and better than recent prior work.  If relatively small amounts of task-specific paired data are available in few-shot scenarios, the Wav2Prompt-LLM combination can be end-to-end (E2E) fine-tuned. The Wav2Prompt-LLM combination then yields greatly improved results relative to an ASR-LLM cascade for the above tasks.
For instance, for English-French ST with the BLOOMZ-7B1 LLM, a Wav2Prompt-LLM combination gave a 8.5 BLEU point increase over an ASR-LLM cascade.

\end{abstract}

\section{Introduction}


Text-based large language models (LLMs) \citep{brown2020language,touvron2023llama, touvron2023llama2, ouyang2022training, le2023bloom} have achieved remarkable performance in a wide range of natural language processing (NLP) tasks \citep{achiam2023gpt}. LLMs are trained on huge quantities of text and are highly flexible. They are able to be applied to a range of tasks that they have not been explicitly trained for, known as the LLM emergent abilities \citep{DBLP:journals/tmlr/WeiTBRZBYBZMCHVLDF22, tang2024salmonn}.
To further expand the use-cases of LLMs, it is important to enable LLMs to handle other modalities including spoken input.

The conventional approach is to use an automatic speech recognition (ASR) model to transcribe speech into text, which is then used as the LLM input. However, this cascaded system suffers from error accumulation and loses valuable acoustic information 
\citep{fathullah2024audiochatllama}. Many studies have explored connecting LLMs directly to the speech acoustic encoder (Encoder-LLM) for various speech tasks such as ASR or speech translation (ST) \citep{10447605, 10389705, 10445874, chen2023x}. However, these approaches restrict the system to a specific task, thereby losing the ability of LLMs to handle a wide range of zero-shot spoken language tasks. Recent work has begun to explore ways to restore the zero-shot capabilities of LLMs, including efforts on audio or speech question-answering (QA) data \cite{gong2024listen, fathullah2024audiochatllama} and the addition of extra steps such as instruction tuning and activation tuning in the training pipeline \cite{tang2024salmonn}. However, these approaches greatly complicate model training.


This paper proposes Wav2Prompt which allows straightforward integration between spoken input and an off-the-shelf  text-based LLM. Wav2Prompt uses a simple training process with the same data as used to train an ASR model.  After training, Wav2Prompt generates representations from speech and uses them as LLM prompts for downstream tasks. It allows the Wav2Prompt-LLM combination to work well in a range of zero-shot spoken language tasks.
However, Wav2Prompt can also give much improved performance when task-specific spoken language data is available in few-shot scenarios via end-to-end (E2E) fine-tuning of Wav2Prompt, without updating the LLM.





Wav2Prompt takes the LLM token embeddings as the training targets to naturally maintain the zero-shot capability of text-based LLMs. This is a key difference to the conventional ASR task, which takes discrete text tokens as the only target and thus leads to the ASR-LLM cascade approach.
However, learning LLM token embeddings is challenging for speech models given the 
difference in input sequence
length between speech and text. Wav2Prompt addresses this issue using a continuous integrate-and-fire (CIF) \cite{9054250} mechanism to generate a label-level speech representation and mean squared error (MSE) can be used to enforce consistency with the LLM token embeddings. Wav2Prompt can therefore be combined with LLM not only for zero-shot speech tasks, but also for E2E fine-tuning in few-shot scenarios, which is a key advantage compared to an ASR-LLM cascade.

Experiments were conducted to evaluate Wav2Prompt on diverse spoken language tasks including speech translation (ST), spoken language understanding (SLU), speech question answering (SQA), and spoken-query-based QA (SQQA), all of these are unseen during training as Wav2Prompt only uses ASR data for training. The results show that the Wav2Prompt-LLM combination could achieve similar performance to the ASR-LLM cascade in zero-shot cases and greatly surpasses the existing Encoder-LLM method \cite{10447605}. In few-shot scenarios, after utilising the limited task-specific available data for E2E fine-tuning, Wav2Prompt showed improved performance over the ASR-LLM cascade for all of these tasks.

The main contributions of this paper can be summarised in three main parts:
\begin{itemize}
    \item Wav2Prompt is proposed, which is, to the best of our knowledge, the first step towards extending LLMs to a range of zero-shot spoken language tasks using only ASR data.
    \item  Task over-fitting to training data is a key issue addressed by Wav2Prompt which has previously limited the application of acoustic encoder enabled LLMs  to other spoken language tasks in a zero-shot fashion  \cite{tang2024salmonn}.  This issue is thoroughly analysed and it is shown that the key step to unlock zero-shot capability is learning LLM token embeddings.
    \item Wav2Prompt achieves similar performance to an ASR-LLM cascade in a range of zero-shot speech tasks. In few-shot scenarios, Wav2Prompt greatly surpasses the performance of an ASR-LLM cascade by leveraging the advantage of E2E fine-tuning.
\end{itemize}

The rest of the paper is organised as follows. Section~\ref{related} reviews related work and discusses  differences to Wav2Prompt. Section~\ref{rethink} analyses the task over-fitting issue.
Section~\ref{method} describes Wav2Prompt in detail. Section~\ref{setup} and ~\ref{results} detail the experimental setup and results. Finally, Section~\ref{conclusion} concludes.

\section{Related Work}
\label{related}
\paragraph{Text-based Large Language Models}
The evolution of text-based LLMs, exemplified by a large increase in model parameters and training data seen in GPT-3 \citep{brown2020language} and PaLM \citep{chowdhery2023palm}, has revolutionised NLP tasks.
This progress has facilitated the development of advanced models such as GPT-4 \citep{achiam2023gpt}, showcasing the remarkable capabilities of LLMs in various domains. Alongside these advancements, "smaller" LLMs like LLaMa \citep{touvron2023llama, touvron2023llama2} have been introduced, achieving a better balance between performance and computational resources. There are several variant models like Vicuna \citep{zhang-etal-2023-sgp} developed from conversation-based fine-tuning and multi-lingual LLM, i.e. BLOOM \citep{le2023bloom}.
One key aspect of LLMs is that they can exhibit remarkable performance in a range of tasks on which they have never been explicitly trained. Examples include zero-shot task transfer \cite{radford2021learning} and few-shot learning \citep{brown2020language}. This is sometimes known as the emergent abilities of LLMs \cite{tang2024salmonn, ma2023investigating}.

\paragraph{Speech-Enabled Large Language Models} 
While discrete self-supervised representations have been explored to build spoken generative LMs \citep{borsos2023audiolm, wang2023neural}, this paper focuses on utilising off-the-shelf text-based LLMs. Recently, several studies have worked on building speech-enabled LLMs to support direct speech input \citep{10447605, 10389705, 10445874, chen2023x, huang2024dynamic}. Since the speech input sequence is normally much longer than the corresponding text, different strategies have been studied for down-sampling. \cite{10447605, 10445874, ma2024embarrassingly} stacks the acoustic encoder output to achieve fixed-rate reduction. \cite{chen2023x} treats multiple modalities as foreign languages, in which CIF is used to obtain the speech representation. 
\cite{10445874} explored the use of Q-Former \citep{li2023blip} which transforms input sequences of varying lengths into fixed-length outputs. However, the emergent abilities of LLM are lost in this task-specific training. Unlike work such as \cite{chen2023x}, which focus on enabling LLMs to handle multi-modal inputs including speech, this paper focuses on maintaining the zero-shot abilities of LLMs.
Some recent work explored regaining the zero-shot abilities of LLMs. In \cite{tang2024salmonn}, an instruction tuning stage followed by an activation tuning stage was introduced after pre-training to alleviate task over-fitting.
In addition, \cite{fathullah2024audiochatllama} simulated a speech QA dataset from ASR data, finding that a speech-enabled LLM trained on it could handle spoken QA tasks and potentially other tasks like ST. Prior work extended speech capabilities for LLM, but they come at the cost of increased complexity and extensive resources based on full-parameter or parameter-efficient training. Since text-based LLMs have a fundamental connection with speech, Wav2Prompt focuses on training a model that can be combined and E2E fine-tuned with a text-based LLM through a simple process while keeping the LLM fixed.

\paragraph{Prompt Tuning}
Fine-tuning LLMs can be expensive. As an alternative, the prompting technique fixes all LLM parameters and uses a prompt to query LLMs for diverse tasks \citep{liu-etal-2022-p}. Early prompting uses simple keyword-based inputs or fill-in-the-blank style prompts \citep{gao-etal-2021-making, Schick2020ExploitingCF}.
For generative LLM, natural language prompts can be used \citep{victor2022multitask, brown2020language}.
However, these discrete prompts can result in sub-optimal performance in numerous cases \citep{autoprompt:emnlp20, liu-etal-2022-p}. Instead, prompt tuning adds trainable continuous embeddings, i.e. continuous prompts, to the original input token embedding \citep{liu2023gpt, Lester2021ThePO}. During training, only the parameters of the continuous prompts are updated \citep{liu-etal-2022-p}. 
This paper follows the prompt tuning, updating only the parameters of Wav2Prompt while keeping LLM fixed in the few-shot scenario.

\section{Analysis of task over-fitting}
\label{rethink}
Task over-fitting~\citep{tang2024salmonn} occurs when the speech-enabled LLM can only perform tasks that are seen during supervised training and shows limited performance on unseen tasks.
This section provides a detailed analysis of this issue which leads to the proposed Wav2Prompt method.

To connect a decoder-only LLM with speech input, speech representations can be prepended to the original
text token embedding sequence and the LLM will be conditioned on these speech representations when predicting the next token in order to perform speech tasks.
To be more specific, in the normal case with a text-based user-input prompt supervision, the next token probabilities of LLM can be formulated as:
\begin{equation}
    p_{n} = p(y_n|\bm{Y}_{0:n-1}, \textbf{P}, \Gamma) \label{normal}
\end{equation}
where $\bm{Y}_{0:n-1}=([sos], y_1, ..., y_{n-1})$ is the sequence of previously predicted tokens and $y_n$ denotes the $n$-th token. $\textbf{P} = (\bm{p}_{1}, \cdots, \bm{p}_{m})$ is the text embedding sequence obtained by feeding the text-based user-input prompt into an LLM embedding layer. $\Gamma$ denotes a task-specific prompt template that contains instructions. When a speech representation $\textbf{S} = (\bm{s}_{1}, \cdots, \bm{s}_{t})$ is prepended to the text-based input as the prompt supervision to replace $\textbf{P}$, Eq.~\ref{normal} can be re-written as $\hat{p}_{n} = p(y_n|\bm{Y}_{0:n-1}, \textbf{S}, \Gamma)$.

It is counter-intuitive to let a fixed text-based LLM attend to speech representation $\textbf{S}$ as it has never seen the speech input at pre-training.
However, after E2E training on supervised data, prior work has shown connecting a fixed LLM with the acoustic encoder (referred to as {Encoder-LLM}) can perform ASR tasks \citep{10447605, 10445874}. 
Since the main building block of LLMs is the attention mechanism which is also the first module that interacts with inputs, the process is simplified to a single attention function ${\rm Att}(Q, K, V)$ for theoretical analysis where the conclusions can be generalised to the entire LLM.
In the normal case with text-based input, suppose the token embedding of $y_{n-1}$ after the LLM embedding layer is $\bm{z}_{n-1}$ and $\textbf{Z} = (\bm{z}_{1}, \cdots, \bm{z}_{n-1})$,
Eq.~\ref{normal} can be expressed as:
\begin{equation}
    \bm{l}_{n} = {\rm Att}(\bm{z}_{n-1},[\textbf{Z}; \textbf{P}; \Gamma], [\textbf{Z}; \textbf{P}; \Gamma]) \label{att}
\end{equation}
where $\bm{l}_{n}$ is the output logits for $p_{n}$, $\bm{z}_{n-1}$ is query, and $[\textbf{Z}; \textbf{P}; \Gamma]$ is keys and values. A speech-enabled LLM must replace $\textbf{P}$ in Eq.~\ref{att} with the speech representation $\textbf{S}$ and make the resulting prediction ${\bm{\hat{l}}_{n}} = {\rm Att}(\bm{z}_{n-1},[\textbf{Z}; \textbf{S}; \Gamma], [\textbf{Z}; \textbf{S}; \Gamma])$ close to ground truth. During E2E supervised training, 
the cross-entropy loss supervises ${\bm{\hat{l}}_{n}}$ while updating $\textbf{S}$.
Even if the lengths and features of $\textbf{S}$ and $\textbf{P}$ are different, $\bm{l}_{n}={\bm{\hat{l}}_{n}}$ is still possible to achieve. Since the attention mechanism is a weighted sum, as long as the weighted sum corresponding to the use of $\textbf{S}$ and $\textbf{P}$ are consistent, correct predictions can be obtained.

However, $z_{n-1}$ and $\Gamma$ differ in different tasks, and $\textbf{S}$ learned on a certain task cannot always guarantee that $\bm{l}_{n}={\bm{\hat{l}}_{n}}$ will still hold with difference $z_{n-1}$ and $\Gamma$ for other tasks, leading to so-called task over-fitting. For example, preliminary experiments found that the Encoder-LLM-based ASR system has trouble following new instructions to perform zero-shot ST.

Prior works relied on task-specific E2E training to implicitly optimise $\textbf{S}$ \citep{10447605, 10445874, chen2023x}, and through complex training pipelines to gradually enable $\textbf{S}$ learning a correct alignment that can preserve the LLM zero-shot capability \cite{tang2024salmonn}. However, to learn a correct $\textbf{S}$, there is already a feasible and clear target, which is the LLM token embedding given that LLM can always flexibly handle text inputs for zero-shot tasks. Unlike prior work,
this paper proposes using LLM embeddings as the target to explicitly guide the learning of the speech representation $\textbf{S}$, which greatly simplifies the process and only requires ASR data for training.

\section{Wav2Prompt}
\label{method}
This paper proposes Wav2Prompt, a model that naturally enables text-based LLM to handle speech input while maintaining the zero-shot capabilities of the original LLM. Wav2Prompt provides a straightforward process that can be combined with LLM using only ASR data for training and does not require subsequent multi-stage tuning, it can serve as an E2E alternative approach that is superior to the conventional ASR-LLM cascade. 

\subsection{Wav2Prompt architecture}
\begin{figure}[t]
    \centering
    \includegraphics[width=138mm]{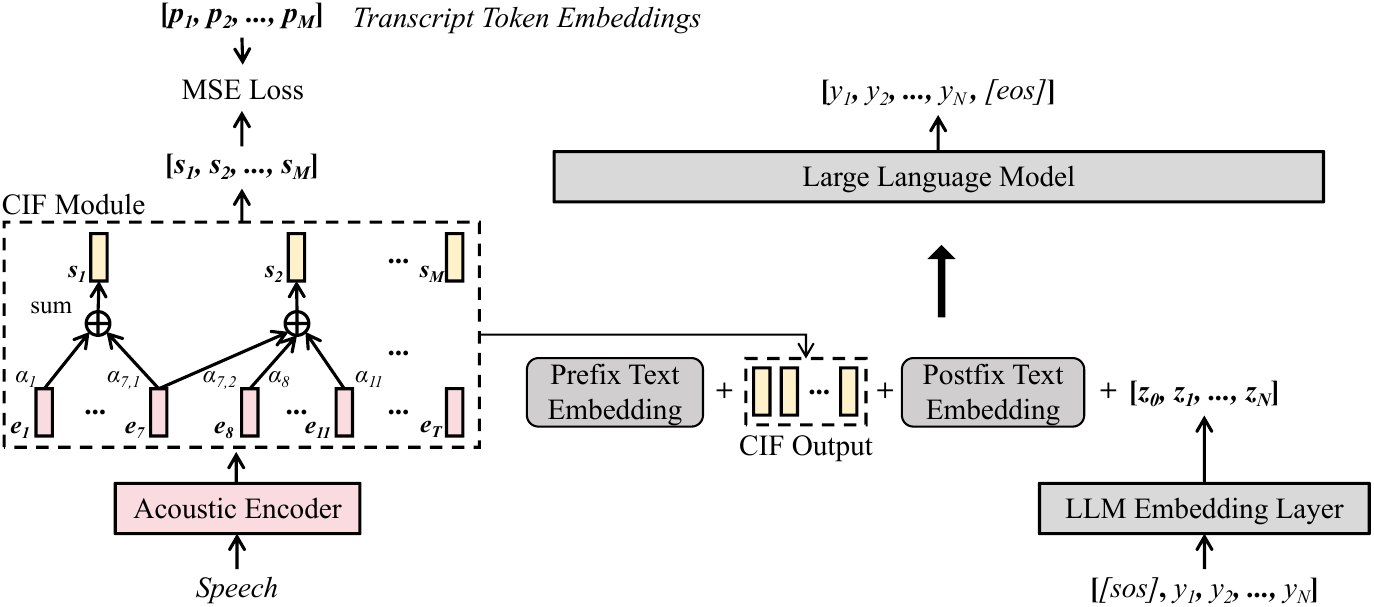}
    \caption{Illustration of the proposed Wav2Prompt architecture. $\bm{\oplus}$ denotes addition. Prefix and postfix text are task-specific prompt templates that can contain instructions. Their embeddings are obtained through the LLM embedding layer, and the transcript token embeddings are the same. } 
    \label{fig:w2p}
\end{figure}

The proposed Wav2Prompt is illustrated in Fig.~\ref{fig:w2p}, which contains three main components: an acoustic encoder, a CIF module, and an LLM (including the LLM embedding layer).
The acoustic encoder and the CIF module extract a label-level speech representation $\textbf{S} = (\bm{s}_{1}, \cdots, \bm{s}_{M})$, which have the same length as the transcript text token embeddings so that mean squared error (MSE) loss can be used to enforce the representation consistency between them. This is one of the main differences from prior work \citep{10447605, 10389705, 10445874, chen2023x, tang2024salmonn, ma2024embarrassingly}, which instead simply down-sampled the acoustic encoder output before feeding it into the LLM.

In this paper, LLM refers to the text-based decoder-only LLM, such as LLaMA \citep{touvron2023llama, touvron2023llama2}. The output of the CIF module is used as the prompt for LLM, while the parameters of LLM are always kept fixed (shown in grey in Fig.~\ref{fig:w2p}) following prompt tuning \citep{liu-etal-2022-p}.

The acoustic encoder employs a Conformer \citep{gulati20_interspeech} structure. Denote the Conformer-based encoder output as 
$\textbf{E} = (\bm{e}_{1}, \cdots, \bm{e}_{T})$, where $T$ is the frame length and is normally much larger than the corresponding text sequence length. To learn a label-level speech representation with a flat-start, i.e. not relying on readily available alignment, the CIF mechanism \cite{9054250} is used. As shown in Fig.~\ref{fig:w2p}, a scalar weight $\alpha_t$ is learnt for each encoder output frame $\bm{e}_t$ and a label-level representation is obtained via weighted addition. Following \citep{9398531, deng2023label}, this paper uses the last dimension of $\bm{e}_{t}$ as the raw scalar attention value of $\alpha_t$ to avoid additional parameters: $\alpha_t = {\rm sigmoid}({e_{t,d}})$ where $d$ is the dimension size of $\bm{e}_{t}$. The weights $\alpha_t$ are accumulated from left to right (i.e. forward along time) until the accumulation exceeds a threshold of 1.0. 
Once reached, the current weight $\alpha_t$ is divided into two parts: one part ensures the current accumulated weight is exactly 1.0, while the remainder is used for the next integration. An example is shown in Fig.~\ref{fig:w2p}, where the threshold 1.0 is achieved when $t=7$ and $\alpha_7$ is divided into $\alpha_{7,1}$ and $\alpha_{7,2}$. The first label-level speech representation $\bm{s}_{1}$ is obtained via:
\begin{equation}
    \bm{s}_{1} = {\rm FC}(\alpha_{1}\cdot\bm{e}_{1,1:d-1} + \cdots + \alpha_{6}\cdot\bm{e}_{6,1:d-1} + \alpha_{7,1}\cdot\bm{e}_{7,1:d-1})  \label{ss}
\end{equation}
where {\rm FC} represents a fully connected layer that maps $\bm{e}_{t,1:d-1}$ to the
LLM embedding dimension. The accumulation is then reset to zero and continues to the right to calculate $\bm{s}_{2}$, $\bm{s}_{3}$, etc. until the end of the encoder output. To ensure that the extracted label-level speech representation sequence $\textbf{S}$ has the exactly same length $M$ as the corresponding transcript token sequence at training, a scaled weight $\hat{\alpha}_t$=$\alpha_t \cdot (M/\sum_{i=1}^T\alpha_i)$ is computed and used to extract $\textbf{S} = (\bm{s}_{1}, \cdots, \bm{s}_{M})$ instead $\alpha_t$ during training. To learn the CIF alignment, 
a quantity loss~\cite{9054250}, $\mathcal{L}_{\rm qua}=|\sum_{i=1}^T\alpha_i - M|$, is computed during training to encourage the accumulated weights approaching the right length $M$.

The label-level speech representation from CIF is then fed into the LLM as a prompt along with task-specific prompt templates that can contain instructions, denoted as prefix and postfix text in Fig.~\ref{fig:w2p}. Suppose $\textbf{emb}^{\rm pre}$ and $\textbf{emb}^{\rm post}$ are the embeddings of the prefix and postfix text sequence, the LLM output logits $\textbf{L} = (\bm{l}_{1}, \cdots, \bm{l}_{N})$ is computed as:
\begin{equation}
    \textbf{L} = {\rm LLM}({\rm Concat}(\textbf{emb}^{\rm pre}, \textbf{S}, \textbf{emb}^{\rm post}, \textbf{Z}))\label{llm-func}
\end{equation}
where $\textbf{Z} = (\bm{z}_{0}, \bm{z}_{1}, \cdots, \bm{z}_{N})$ is the embeddings of LLM input $([sos], y_1, ..., y_{N})$ as shown in Fig.~\ref{fig:w2p}, Concat$(\cdot)$ denotes concatenation of vector sequences, and ${\rm LLM}(\cdot)$ denotes the LLM function that takes the concatenated vector sequence as inputs and outputs logits $\mathbf{L}$.


\subsection{Training}
Wav2Prompt is trained  using only ASR data. First, the scaled weight $\hat{\alpha}_t$ is used in training to extract the label-level speech representation $\textbf{S} = (\bm{s}_{1}, \cdots, \bm{s}_{M})$ that has the same length as transcript token sequence. After feeding the transcript text tokens into the LLM embedding layer, the embedding sequence $\textbf{P} = (\bm{p}_{1}, \cdots, \bm{p}_{M})$ is used as the training target of $\textbf{S}$, and an MSE loss is computed:
\begin{equation}
    \mathcal{L}_{\rm MSE} = \sum_{m=1}^{M} {\rm MSE}(\bm{s}_{m},\bm{p}_{m})
\end{equation}
In addition, $\textbf{P}$ is fed into the LLM and
a cross-entropy (CE) loss $\mathcal{L}_{\rm CE}$ is computed between the LLM output logits $\textbf{L}$ and target transcripts, ensuring that the obtained $\textbf{S}$ can be understood by the fixed LLM.
Finally, the quantity loss $\mathcal{L}_{\rm qua}$ is also included to learn the CIF alignment as mentioned above.
Therefore, the overall training objective $\mathcal{L}_{\rm Train}$ of Wav2Prompt is: 
 \begin{equation}
    \mathcal{L}_{\rm Train}=\mathcal{L}_{\rm CE} + \gamma \mathcal{L}_{\rm MSE}+\mu \mathcal{L}_{\rm qua} \label{obj-train}
\end{equation}
where $\gamma$ and $\mu$ are hyper-parameters.

\subsection{Application for zero-shot and few-shot tasks}
Wav2Prompt is only trained on the ASR task, but it can be combined with the text-based LLM (denoted as Wav2Prompt-LLM) for other speech tasks in zero-shot and few-shot fashions. In this paper, zero-shot refers to not using any task-specific paired data for fine-tuning. For example, for the speech translation (ST) task, zero-shot refers to not fine-tuning the Wav2prompt-LLM system using paired source-language speech and target-language text data. Few-shot in this paper refers to using a very limited amount of paired data for fine-tuning.

\paragraph{Zero-shot} Wav2Prompt preserves the flexible zero-shot capabilities of LLMs. With ASR-data trained Wav2Prompt, the generated label-level speech representation $\textbf{S}$ is fed into the LLM as a prompt, and only the instructions in the prefix and postfix text (as in Fig.~\ref{fig:w2p}) need to be modified for unseen tasks. For example, to perform the ST task, the postfix text becomes ``Translate the English text into French". The original weight ${\alpha}_t$ is used instead of $\hat{\alpha}_t$ as the transcript length is unknown during inference. There is essentially no difference from a normal text-based LLM for different text tasks.

\paragraph{Few-shot} Compared to an ASR-LLM cascade, an important advantage of Wav2Prompt is that it can be E2E combined with an LLM so that paired data can be used to fine-tune in an E2E fashion. In the few-shot case, the prefix and postfix text are modified as in the zero-shot case. 
In addition, during E2E fine-tuning, in order to simplify the process, the MSE loss is not used as it requires high-quality ASR transcription and the goal here is to learn the downstream task rather than to keep the zero-shot ability. Another benefit is that the length of the speech representation $\textbf{S}$ does not need to exactly match that of its corresponding transcript, allowing ${\alpha}_t$ to be used to match the inference condition.
This overcomes the mismatch between training and inference in the original CIF (i.e., ${\alpha}_t$ was used during training while $\hat{\alpha}_t$ was used during inference), which was solved after the few-shot fine-tuning stage.
Finally, the quantity loss $\mathcal{L}_{\rm qua}$ is still computed, because preliminary experiments have shown that without the regularising effect of $\mathcal{L}_{\rm qua}$, the length of $\textbf{S}$ can undergo drastic changes during optimisation, hindering convergence.
Hence, the fine-tuning objective $\mathcal{L}_{\rm tune}$ of Wav2Prompt is: 
 \begin{equation}
    \mathcal{L}_{\rm tune}= \mathcal{L}_{\rm CE}+\mu \mathcal{L}_{\rm qua} \label{obj}
\end{equation}
where hyper-parameter $\mu$ keeps the same as Eq.~\ref{obj-train} for simplicity.

\section{Experimental setup}
\label{setup}

After training on ASR data, Wav2Prompt was evaluated on a range of unseen tasks, including speech translation (ST), spoken language understanding (SLU), speech question answering (SQA), and spoken-query-based QA (SQQA) tasks. For SQA, the system needs to answer text-based questions according to the content of the speech, while for SQQA, it needs to answer the spoken question.

\subsection{Datasets}
ST experiments were conducted on Europarl-ST \citep{jairsan2020a} English-Spanish (En-ES) and English-French (En-Fr) pairs. The corresponding English ASR data was used to train Wav2Prompt and the ASR models. For the few-shot scenarios, 10-hour paired data was randomly selected from the training data set as limited fine-tuning data.


For SLU, SQA, and SQQA tasks, the LibriSpeech corpus \citep{7178964} was used as the ASR data. For SLU, the Fluent Speech Commands (FSC) corpus \cite{lugosch19_interspeech} was used to conduct the intent classification task. In the few-shot case, 2 hours of paired data were randomly selected from the training data set as limited fine-tuning data. For the SQA task, the question-answer pairs provided by \citep{tang2024salmonn} were used to augment the dev-clean set. For the SQQA task, the WikiQA \citep{yang-etal-2015-wikiqa} test set with synthesised speech queries provided by \citep{tang2024salmonn} was used.
More details of the datasets used are listed in Appendix~\ref{stat}.

\subsection{Model specifications}
Four different systems were built to compare with Wav2Prompt, and all these models used a 12-layer Conformer encoder. 
The LLMs used in this paper were always fixed following
prompt tuning.


\paragraph{Wav2Prompt-LLM} Based on the Conformer encoder, Wav2Prompt only used an extra fully connected (FC) layer that mapped the speech representation to the LLM embedding dimension (i.e. 4096) as mentioned in Eq.~\ref{ss}. $\gamma$ and $\mu$ in Eq.~\ref{obj-train} were set to 20 and 0.05, respectively. The LLM was fixed.

\paragraph{ASR-LLM Cascade} Based on the Conformer encoder, a connectionist temporal classification \citep{graves2006connectionist} (CTC)-based ASR model was built and only had an extra FC output layer that mapped the encoder output to the vocabulary size (1000) and BPE \citep{gage1994new} modelling units were used. The recognised text from the ASR model was fed into the LLM, along with the prefix and postfix text, forming a cascaded system. In this paper, the proposed Wav2Prompt aims to achieve similar results to the ASR-LLM Cascade in the zero-shot scenarios.

\paragraph{Oracle-LLM} An oracle system was built, in which the speech ground truth transcripts were fed into the LLM. The prefix and postfix text remained the same as the ASR-LLM Cascade system.

\paragraph{Encoder-LLM} A prior speech-enabled LLM method \citep{10447605} was implemented. Based on the Conformer encoder, every 8 consecutive encoder output frames were stacked to down-sample the sequence length. Then, an extra FC layer was used to map the stacked encoder output to the LLM embedding dimension (i.e. 4096) before being fed into the LLM just like Wav2Prompt. Following \citep{10447605},  the trained CTC-based ASR model was used to initialise the encoder parameters before the Encoder-LLM was E2E trained on ASR data. In this paper, the proposed Wav2Prompt aims to greatly surpass the Encoder-LLM in the zero-shot scenarios.

\paragraph{Flat-start Encoder-LLM} This paper further explores directly training the Encoder-LLM on unseen tasks via few-shot learning in an E2E fashion without training on ASR data.
This is to evaluate the importance of the alignment learned during ASR training to other tasks. This system is referred to as Flat-start Encoder-LLM in this paper. The encoder was still initialised by the CTC-based ASR model.

Therefore, the trainable component of all these built systems (except for the Oracle-LLM) consisted of the same encoder along with an additional FC layer. More details and the task-specific prompt templates used can be found in Appendix~\ref{hyper-detail} and \ref{fixed}.

\subsection{LLMs and metrics}
For the ST task, LLaMA-2-7B was used. To evaluate the proposed Wav2Prompt on an LLM more proficient in translation, BLOOMZ-7B1 \citep{muennighoff2022crosslingual} was also employed.
Case-sensitive detokenised BLEU \citep{papineni2002bleu} results are reported to evaluate translation quality.

For the SLU, SQA, and SQQA tasks, Vicuna-7B-1.5 \citep{zhang-etal-2023-sgp} was used, and ASR performance was also evaluated.
Vicuna-7B-1.5 is a fine-tuned version of LLaMA-2 and adapted specifically to chat applications. Therefore, it shows good performance in following user instructions and is suitable for SLU and QA tasks that require an understanding capability. The word error rate (WER) was used to evaluate the ASR quality. For the SLU task, accuracy was used to measure the intent classification.
For the SQA and SQQA task, following \cite{Maaz2023VideoChatGPT}, Mistral-7B-Instruct-v0.2, which is an instruction fine-tuned LLM, was used to evaluate whether the answers predicted were correct based on the question and the right answer. Accuracy was used as the metric.
The prompt template used to measure the accuracy was listed in Appendix~\ref{qq-eval}. 

\section{Experimental results}
\label{results}
\subsection{ST task results}

\begin{table}[ht]
  \caption{BLEU ($\uparrow$) results on the test sets of Europarl-ST En-Es and En-Fr pairs with different LLM. Zero-shot means no speech-translation paired data available for fine-tuning, while few-shot means 10-hour speech-translation paired data available}
  \label{tab:st}
  \centering
  \setlength{\tabcolsep}{1.5mm}
  \renewcommand\arraystretch{1.1}
  \begin{tabular}{l c | c c| c  c}
    \Xhline{3\arrayrulewidth}
    \multirow{2}{*}{Model}&\multirow{2}{*}{LLM}&\multicolumn{2}{c|}{En-Es} &\multicolumn{2}{c}{En-Fr}\\
    &&Zero-shot&Few-shot&Zero-shot&Few-shot\\
    \hline
    Oracle-LLM &LLaMA-2-7B&26.1&26.1&18.8&18.8\\
    \hline
    ASR-LLM Cascade&LLaMA-2-7B&14.0&14.0&10.4&10.4\\
    Encoder-LLM&LLaMA-2-7B&6.2&19.1&4.0&16.3\\
    Flat-start Encoder-LLM&LLaMA-2-7B&---&1.8&---&2.6\\
    Proposed Wav2Prompt-LLM&LLaMA-2-7B&\textbf{13.8}&\textbf{19.9}&\textbf{9.2}&\textbf{16.9}\\
    \hline
    Oracle-LLM &BLOOMZ-7B1&32.9&32.9&24.4&24.4\\
    \hline
    ASR-LLM Cascade&BLOOMZ-7B1&17.7&17.7&15.1&15.1\\
    Encoder-LLM&BLOOMZ-7B1&7.9&25.8&5.5&23.0\\
    Flat-start Encoder-LLM&BLOOMZ-7B1&---&0.7&---&0.5\\
    Proposed Wav2Prompt-LLM&BLOOMZ-7B1&\textbf{18.3}&\textbf{26.0}&\textbf{12.7}&\textbf{23.6}\\
    \Xhline{3\arrayrulewidth}
  \end{tabular}
\end{table}

The ST results are shown in Table~\ref{tab:st}. Since the Oracle-LLM and ASR-LLM Cascade systems cannot utilise speech-to-translation paired data for E2E fine-tuning in the few-shot scenario, their few-shot results are the same as their zero-shot results. Note this paper uses prompt tuning \citep{liu-etal-2022-p} and the LLM was fixed. As shown in Table~\ref{tab:st}, Wav2Prompt-LLM achieves competitive results in the zero-shot scenario compared to ASR-LLM Cascade, which shows that the proposed Wav2Prompt maintains the advantage of the LLM zero-shot capability. Moreover, Wav2Prompt-LLM greatly outperformed the Encoder-LLM in the zero-shot scenario, which is in line with the findings of the previous work \citep{tang2024salmonn} that Encoder-LLM overfits to the ASR task when trained on ASR data, and exhibited limited 
performance in unseen ST tasks.

In the few-shot scenario, after fine-tuning with limited paired data, Wav2Prompt-LLM exceeded the performance of the ASR-LLM Cascade, which is the main advantage of Wav2Prompt compared to standard ASR models combined with LLMs. In addition, after fine-tuning with limited data, the Encoder-LLM also shows strong results, only slightly poorer than Wav2Prompt. This indicates that although the Encoder-LLM overfits to the ASR tasks, this issue can be ameliorated via few-shot fine-tuning.
The Flat-start Encoder-LLM results show that the limited data in the few-shot case was insufficient for an encoder, with no exposure to the LLM, to learn how to connect to the LLM.

While BLOOMZ-7B1 gives higher BLEU scores than LLaMA-2-7B in translation tasks, the experimental conclusions are consistent across BLOOMZ-7B1 and LLaMA-2-7B, and it is also consistent across the En-Es and En-Fr tasks. Wav2Prompt can achieve similar results to the ASR-LLM Cascade in zero-shot scenarios and greatly outperforms the Encoder-LLM. In the few-shot scenarios, Wav2Prompt can surpasses  the ASR-LLM Cascade. In conclusion, Wav2Prompt is an effective E2E alternative to conventional ASR when combined with an LLM.

\subsection{SLU task results}

\begin{table}[ht]
  \caption{Cross-domain intent classification accuracy (\%) ($\uparrow$) on FSC corpus for models trained on LibriSpeech corpus. In the few-shot, 2-hour paired data is available for fine-tuning.}
  \label{tab:slu}
  \centering
  \setlength{\tabcolsep}{1.5mm}
  \renewcommand\arraystretch{1.1}
  \begin{tabular}{l c| c |c}
    \Xhline{3\arrayrulewidth}
    {Model}&{LLM}&Zero-shot&Few-shot\\
    \hline
    Oracle-LLM &Vicuna-7B-1.5&97.86\%&97.86\%\\
    \hline
    ASR-LLM Cascade&Vicuna-7B-1.5&75.80\%&93.22\%\\
    Encoder-LLM&Vicuna-7B-1.5&30.00\%&97.57\%\\
    Proposed Wav2Prompt-LLM&Vicuna-7B-1.5&\textbf{71.55\%}&\textbf{98.10}\%\\
    \Xhline{3\arrayrulewidth}
  \end{tabular}
\end{table}

The SLU task requires the LLM to understand the content of speech, and in this section, to be able to classify the intent of the speech. Furthermore, this experiment was conducted in a cross-domain scenario, i.e., the model trained on LibriSpeech ASR data was directly evaluated on FSC data. The results of intent classification are presented in Table~\ref{tab:slu}. In the few-shot scenario, although the ASR-LLM Cascade system cannot utilise SLU paired data for E2E fine-tuning, considering the cross-domain scenario, the limited ASR data corresponding to SLU paired data can be used by the ASR model for fine-tuning to achieve domain adaptation. Therefore, the few-shot results of the ASR-LLM Cascade showed a noticeable improvement compared to the zero-shot results. In the zero-shot scenario, Wav2Prompt-LLM achieved performance close to the ASR-LLM Cascade and greatly surpassed the Encoder-LLM, which again shows that the Encoder-LLM overfits to the ASR task that it was used in training, while Wav2Prompt retains the LLM zero-shot ability. In the few-shot scenario, even compared to the domain-adapted ASR-LLM Cascade, Wav2Prompt-LLM gave an improved performance, demonstrating the advantage of Wav2Prompt-LLM in E2E fine-tuning.

\subsection{SQA and SQQA task results}

\begin{table}[ht]
    \caption{Accuracy (\%) ($\uparrow$) of zero-shot SQA on LibriSpeech augmented with question-answer pairs and zero-shot SQQA on synthesised WikiQA.}
  \label{tab:qa}
  \centering
  \setlength{\tabcolsep}{1.5mm}
  \renewcommand\arraystretch{1.1}
  \begin{tabular}{l c| c| c }
    \Xhline{3\arrayrulewidth}
    {Model}&{LLM}&SQA&SQQA\\
    \hline
    Oracle-LLM &Vicuna-7B-1.5&78.10\%&68.10\%\\
    \hline
    ASR-LLM Cascade&Vicuna-7B-1.5&73.82\%&51.36\%\\
    Encoder-LLM&Vicuna-7B-1.5&64.83\%& 19.54\%\\
    Proposed Wav2Prompt-LLM&Vicuna-7B-1.5&\textbf{74.21\%}&\textbf{51.58\%}\\
    \Xhline{3\arrayrulewidth}
  \end{tabular}
\end{table}

The zero-shot results of SQA and SQQA are shown in Table~\ref{tab:qa}, where the SQQA task
was evaluated in a cross-domain scenario, making it more challenging.
Overall, the ASR-LLM Cascade and Wav2Prompt-LLM achieved similar results, with Wav2Prompt-LLM being slightly better. 
Compared to these two systems, the Encoder-LLM showed a noticeable performance gap, especially on the SQQA task, again due to task over-fitting.
Hence, the experimental conclusion is consistent with the above ST and SLU experiments that Wav2Prompt effectively retains the LLM zero-shot ability, making it an E2E alternative to conventional ASR when combined with an LLM.

\subsection{ASR task results}

\begin{table}[ht]
  \caption{ASR WER ($\downarrow$) results on LibriSpeech test sets.}
  \label{tab:asr}
  \centering
  \setlength{\tabcolsep}{1.2mm}
  \renewcommand\arraystretch{1.1}
  \begin{tabular}{l c c| c c |c c}
    \Xhline{3\arrayrulewidth}
    \multirow{2}{*}{Model}&\multirow{2}{*}{Encoder}&\multirow{2}{*}{LLM}&\multicolumn{2}{c|}{test}&\multicolumn{2}{c}{dev}\\
    &&&clean&other&clean&other\\
    \hline
    SLAM-ASR \citep{ma2024embarrassingly}&Whisper-small& LLaMA-2-7B& 6.4 & 10.9&---&---\\
    SLAM-ASR \citep{ma2024embarrassingly}&Whisper-small& LLaMA-2-7B-chat&  4.5 &8.9&---&---\\
    SLAM-ASR \citep{ma2024embarrassingly}&Whisper-small& Vicuna-7B-1.5&   4.2& 9.5&---&---\\
    \hline
    Encoder-LLM&Conformer&Vicuna-7B-1.5&4.5&10.8&4.7&11.2\\
    Proposed Wav2Prompt-LLM&Conformer&Vicuna-7B-1.5&\textbf{3.9}&\textbf{8.2}&\textbf{4.0}&\textbf{8.6}\\
    \Xhline{3\arrayrulewidth}
  \end{tabular}
  \begin{tablenotes}
  \footnotesize
  \item{\hspace{-3.5mm}*}{Results with Whisper-small listed as Whisper-small (87M) has similar parameter size to our Conformer (83M). Direct comparison not feasible as Whisper-small trained on a very large dataset (680k hours).}
  \end{tablenotes}
\end{table}

While this paper focuses on downstream tasks such as ST, SLU, SQA, and SQQA, the results of ASR are also given in Table~\ref{tab:asr}. Wav2Prompt-LLM gave competitive ASR performance on the LibriSpeech benchmark data and outperformed the Encoder-LLM, with this improvement being statistically significant ($p$<0.1\%) according to the matched pair sentence segment test \citep{115546}.
This shows that enforcing similarity between speech representations and LLM embeddings is beneficial for the LLM to transcribe speech input.


\subsection{Ablation studies}

\begin{table}[ht]
  \caption{Ablation studies on the MSE loss for Wav2Prompt. Zero-shot BLEU ($\uparrow$) results were shown.}
  \label{tab:ablation}
  \centering
  \setlength{\tabcolsep}{1.5mm}
  \renewcommand\arraystretch{1.1}
  \begin{tabular}{l c| c c }
    \Xhline{3\arrayrulewidth}
    {Model}&{LLM}&En-Es&En-Fr\\
    \hline
    Oracle-LLM &LLaMA-2-7B&26.1&18.8\\
    ASR-LLM Cascade&LLaMA-2-7B&14.0&10.4\\
    Encoder-LLM&LLaMA-2-7B&6.2&4.0\\
    \hline
    Proposed Wav2Prompt-LLM&LLaMA-2-7B&13.8&9.2\\
    Proposed Wav2Prompt-LLM w/o MSE Loss&LLaMA-2-7B&5.4&3.3\\
    \Xhline{3\arrayrulewidth}
  \end{tabular}
\end{table}

In Wav2Prompt, the MSE loss is used to enforce the consistency between label-level speech representation and LLM embeddings, thereby preserving the zero-shot capability of text-based LLMs. In this section, ablation studies were conducted to validate the importance of the MSE loss. As shown in Table~\ref{tab:ablation}, without the MSE loss, the zero-shot ST performance of Wav2Prompt-LLM dropped noticeably, so that Wav2Prompt-LLM overfits to the ASR task, similar to how the Encoder-LLM struggled with unseen tasks. Therefore, learning to match the LLM embeddings is the key to unlocking the zero-shot capability of speech-enabled LLM.


\section{Conclusions}
\label{conclusion}
This paper describes Wav2Prompt, which proposes a method to connect spoken input with text-based LLMs using only ASR training data while retaining the zero-shot capability for other spoken language tasks.
Wav2Prompt extracts label-level speech representations using the CIF mechanism and explicitly enforces the consistency between the speech representations and LLM embeddings using the MSE loss function, thus avoiding the issue of task over-fitting. Experiments on a range of tasks, including ST, SLU, SQA, and SQQA, showed that Wav2Prompt can achieve results close to the ASR-LLM cascade system in zero-shot scenarios and greatly outperforms the existing speech-enabled LLM method. It gives results that exceed those of the ASR-LLM cascade in few-shot scenarios. Wav2Prompt is an E2E trainable alternative to conventional ASR when combined with text LLMs.

\begin{ack}
Keqi Deng is funded by the Cambridge Trust. This work has been performed using resources provided by the Cambridge Tier-2 system operated by the University of Cambridge Research Computing Service (www.hpc.cam.ac.uk) funded by EPSRC Tier-2 capital grant EP/T022159/1.
\end{ack}
\bibliographystyle{elsarticle-harv}
\bibliography{ref}

\begin{thebibliography}{49}
\expandafter\ifx\csname natexlab\endcsname\relax\def\natexlab#1{#1}\fi
\providecommand{\url}[1]{\texttt{#1}}
\providecommand{\href}[2]{#2}
\providecommand{\path}[1]{#1}
\providecommand{\DOIprefix}{doi:}
\providecommand{\ArXivprefix}{arXiv:}
\providecommand{\URLprefix}{URL: }
\providecommand{\Pubmedprefix}{pmid:}
\providecommand{\doi}[1]{\href{http://dx.doi.org/#1}{\path{#1}}}
\providecommand{\Pubmed}[1]{\href{pmid:#1}{\path{#1}}}
\providecommand{\bibinfo}[2]{#2}
\ifx\xfnm\relax \def\xfnm[#1]{\unskip,\space#1}\fi
\bibitem[{Achiam et~al.(2024)Achiam, Adler, Agarwal, Ahmad, Akkaya, Aleman, Almeida, Altenschmidt, Altman, Anadkat et~al.}]{achiam2023gpt}
\bibinfo{author}{Achiam, J.}, \bibinfo{author}{Adler, S.}, \bibinfo{author}{Agarwal, S.}, \bibinfo{author}{Ahmad, L.}, \bibinfo{author}{Akkaya, I.}, \bibinfo{author}{Aleman, F.L.}, \bibinfo{author}{Almeida, D.}, \bibinfo{author}{Altenschmidt, J.}, \bibinfo{author}{Altman, S.}, \bibinfo{author}{Anadkat, S.}, et~al., \bibinfo{year}{2024}.
\newblock \bibinfo{title}{{GPT-4} technical report}.
\newblock \href{http://arxiv.org/abs/2303.08774}{{\tt arXiv:2303.08774}}.
\bibitem[{Borsos et~al.(2023)Borsos, Marinier, Vincent, Kharitonov, Pietquin, Sharifi, Roblek, Teboul, Grangier, Tagliasacchi et~al.}]{borsos2023audiolm}
\bibinfo{author}{Borsos, Z.}, \bibinfo{author}{Marinier, R.}, \bibinfo{author}{Vincent, D.}, \bibinfo{author}{Kharitonov, E.}, \bibinfo{author}{Pietquin, O.}, \bibinfo{author}{Sharifi, M.}, \bibinfo{author}{Roblek, D.}, \bibinfo{author}{Teboul, O.}, \bibinfo{author}{Grangier, D.}, \bibinfo{author}{Tagliasacchi, M.}, et~al., \bibinfo{year}{2023}.
\newblock \bibinfo{title}{{AudioLM}: A language modeling approach to audio generation}.
\newblock \bibinfo{journal}{IEEE/ACM Transactions on Audio, Speech, and Language Processing} \bibinfo{volume}{31}, \bibinfo{pages}{2523--2533}.
\newblock \DOIprefix\doi{10.1109/TASLP.2023.3288409}.
\bibitem[{Brown et~al.(2020)Brown, Mann, Ryder, Subbiah, Kaplan, Dhariwal, Neelakantan, Shyam, Sastry, Askell et~al.}]{brown2020language}
\bibinfo{author}{Brown, T.}, \bibinfo{author}{Mann, B.}, \bibinfo{author}{Ryder, N.}, \bibinfo{author}{Subbiah, M.}, \bibinfo{author}{Kaplan, J.D.}, \bibinfo{author}{Dhariwal, P.}, \bibinfo{author}{Neelakantan, A.}, \bibinfo{author}{Shyam, P.}, \bibinfo{author}{Sastry, G.}, \bibinfo{author}{Askell, A.}, et~al., \bibinfo{year}{2020}.
\newblock \bibinfo{title}{Language models are few-shot learners}, in: \bibinfo{booktitle}{Proc. NeurIPS}.
\newblock \URLprefix \url{https://proceedings.neurips.cc/paper_files/paper/2020/file/1457c0d6bfcb4967418bfb8ac142f64a-Paper.pdf}.
\bibitem[{Chen et~al.(2023)Chen, Han, Zhao, Zhang, Shi, Xu and Xu}]{chen2023x}
\bibinfo{author}{Chen, F.}, \bibinfo{author}{Han, M.}, \bibinfo{author}{Zhao, H.}, \bibinfo{author}{Zhang, Q.}, \bibinfo{author}{Shi, J.}, \bibinfo{author}{Xu, S.}, \bibinfo{author}{Xu, B.}, \bibinfo{year}{2023}.
\newblock \bibinfo{title}{{X-LLM}: Bootstrapping advanced large language models by treating multi-modalities as foreign languages} \href{http://arxiv.org/abs/2305.04160}{{\tt arXiv:2305.04160}}.
\bibitem[{Chowdhery et~al.(2023)Chowdhery, Narang, Devlin, Bosma, Mishra, Roberts, Barham, Chung, Sutton, Gehrmann et~al.}]{chowdhery2023palm}
\bibinfo{author}{Chowdhery, A.}, \bibinfo{author}{Narang, S.}, \bibinfo{author}{Devlin, J.}, \bibinfo{author}{Bosma, M.}, \bibinfo{author}{Mishra, G.}, \bibinfo{author}{Roberts, A.}, \bibinfo{author}{Barham, P.}, \bibinfo{author}{Chung, H.W.}, \bibinfo{author}{Sutton, C.}, \bibinfo{author}{Gehrmann, S.}, et~al., \bibinfo{year}{2023}.
\newblock \bibinfo{title}{{PaLM}: Scaling language modeling with pathways}.
\newblock \bibinfo{journal}{Journal of Machine Learning Research} \bibinfo{volume}{24}, \bibinfo{pages}{1--113}.
\newblock \URLprefix \url{http://jmlr.org/papers/v24/22-1144.html}.
\bibitem[{Deng and Woodland(2023)}]{deng2023label}
\bibinfo{author}{Deng, K.}, \bibinfo{author}{Woodland, P.C.}, \bibinfo{year}{2023}.
\newblock \bibinfo{title}{Label-synchronous neural transducer for adaptable online {E2E} speech recognition}.
\newblock \href{http://arxiv.org/abs/2311.11353}{{\tt arXiv:2311.11353}}.
\bibitem[{Dong and Xu(2020)}]{9054250}
\bibinfo{author}{Dong, L.}, \bibinfo{author}{Xu, B.}, \bibinfo{year}{2020}.
\newblock \bibinfo{title}{{CIF}: Continuous integrate-and-fire for end-to-end speech recognition}, in: \bibinfo{booktitle}{Proc. ICASSP}.
\newblock \DOIprefix\doi{10.1109/ICASSP40776.2020.9054250}.
\bibitem[{Fathullah et~al.(2024a)Fathullah, Wu, Lakomkin, Jia, Shangguan, Li, Guo, Xiong, Mahadeokar, Kalinli, Fuegen and Seltzer}]{10447605}
\bibinfo{author}{Fathullah, Y.}, \bibinfo{author}{Wu, C.}, \bibinfo{author}{Lakomkin, E.}, \bibinfo{author}{Jia, J.}, \bibinfo{author}{Shangguan, Y.}, \bibinfo{author}{Li, K.}, \bibinfo{author}{Guo, J.}, \bibinfo{author}{Xiong, W.}, \bibinfo{author}{Mahadeokar, J.}, \bibinfo{author}{Kalinli, O.}, \bibinfo{author}{Fuegen, C.}, \bibinfo{author}{Seltzer, M.}, \bibinfo{year}{2024}a.
\newblock \bibinfo{title}{Prompting large language models with speech recognition abilities}, in: \bibinfo{booktitle}{Proc. ICASSP}.
\newblock \DOIprefix\doi{10.1109/ICASSP48485.2024.10447605}.
\bibitem[{Fathullah et~al.(2024b)Fathullah, Wu, Lakomkin, Li, Jia, Shangguan, Mahadeokar, Kalinli, Fuegen and Seltzer}]{fathullah2024audiochatllama}
\bibinfo{author}{Fathullah, Y.}, \bibinfo{author}{Wu, C.}, \bibinfo{author}{Lakomkin, E.}, \bibinfo{author}{Li, K.}, \bibinfo{author}{Jia, J.}, \bibinfo{author}{Shangguan, Y.}, \bibinfo{author}{Mahadeokar, J.}, \bibinfo{author}{Kalinli, O.}, \bibinfo{author}{Fuegen, C.}, \bibinfo{author}{Seltzer, M.}, \bibinfo{year}{2024}b.
\newblock \bibinfo{title}{{AudioChatLlama}: Towards general-purpose speech abilities for {LLMs}}.
\newblock \href{http://arxiv.org/abs/2311.06753}{{\tt arXiv:2311.06753}}.
\bibitem[{Gage(1994)}]{gage1994new}
\bibinfo{author}{Gage, P.}, \bibinfo{year}{1994}.
\newblock \bibinfo{title}{A new algorithm for data compression}.
\newblock \bibinfo{journal}{The C Users Journal} \bibinfo{volume}{12}, \bibinfo{pages}{23--38}.
\newblock \URLprefix \url{https://api.semanticscholar.org/CorpusID:59804030}.
\bibitem[{Gao et~al.(2021)Gao, Fisch and Chen}]{gao-etal-2021-making}
\bibinfo{author}{Gao, T.}, \bibinfo{author}{Fisch, A.}, \bibinfo{author}{Chen, D.}, \bibinfo{year}{2021}.
\newblock \bibinfo{title}{Making pre-trained language models better few-shot learners}, in: \bibinfo{editor}{Zong, C.}, \bibinfo{editor}{Xia, F.}, \bibinfo{editor}{Li, W.}, \bibinfo{editor}{Navigli, R.} (Eds.), \bibinfo{booktitle}{Proc. {ACL/IJCNLP} {(1)}}.
\newblock \URLprefix \url{https://aclanthology.org/2021.acl-long.295}.
\bibitem[{Gong et~al.(2024)Gong, Luo, Liu, Karlinsky and Glass}]{gong2024listen}
\bibinfo{author}{Gong, Y.}, \bibinfo{author}{Luo, H.}, \bibinfo{author}{Liu, A.H.}, \bibinfo{author}{Karlinsky, L.}, \bibinfo{author}{Glass, J.R.}, \bibinfo{year}{2024}.
\newblock \bibinfo{title}{Listen, think, and understand}, in: \bibinfo{booktitle}{Proc. ICLR}.
\newblock \URLprefix \url{https://openreview.net/forum?id=nBZBPXdJlC}.
\bibitem[{Graves et~al.(2006)Graves, Fern{\'a}ndez, Gomez and Schmidhuber}]{graves2006connectionist}
\bibinfo{author}{Graves, A.}, \bibinfo{author}{Fern{\'a}ndez, S.}, \bibinfo{author}{Gomez, F.}, \bibinfo{author}{Schmidhuber, J.}, \bibinfo{year}{2006}.
\newblock \bibinfo{title}{Connectionist temporal classification: Labelling unsegmented sequence data with recurrent neural networks}, in: \bibinfo{booktitle}{Proc. {ICML}}.
\newblock \URLprefix \url{https://api.semanticscholar.org/CorpusID:9901844}.
\bibitem[{Gulati et~al.(2020)Gulati, Qin, Chiu, Parmar, Zhang, Yu, Han, Wang, Zhang, Wu and Pang}]{gulati20_interspeech}
\bibinfo{author}{Gulati, A.}, \bibinfo{author}{Qin, J.}, \bibinfo{author}{Chiu, C.C.}, \bibinfo{author}{Parmar, N.}, \bibinfo{author}{Zhang, Y.}, \bibinfo{author}{Yu, J.}, \bibinfo{author}{Han, W.}, \bibinfo{author}{Wang, S.}, \bibinfo{author}{Zhang, Z.}, \bibinfo{author}{Wu, Y.}, \bibinfo{author}{Pang, R.}, \bibinfo{year}{2020}.
\newblock \bibinfo{title}{Conformer: Convolution-augmented {Transformer} for speech recognition}, in: \bibinfo{booktitle}{Proc. Interspeech}.
\newblock \DOIprefix\doi{10.21437/Interspeech.2020-3015}.
\bibitem[{He and Garner(2023)}]{he23_interspeech}
\bibinfo{author}{He, M.}, \bibinfo{author}{Garner, P.N.}, \bibinfo{year}{2023}.
\newblock \bibinfo{title}{Can {ChatGPT} detect intent? evaluating large language models for spoken language understanding}, in: \bibinfo{booktitle}{Proc. Interspeech}.
\newblock \DOIprefix\doi{10.21437/Interspeech.2023-1799}.
\bibitem[{Huang et~al.(2024)Huang, Lu, Wang, Hsiao, Kuan, Wu, Arora, Chang, Shi, Peng et~al.}]{huang2024dynamic}
\bibinfo{author}{Huang, C.y.}, \bibinfo{author}{Lu, K.H.}, \bibinfo{author}{Wang, S.H.}, \bibinfo{author}{Hsiao, C.Y.}, \bibinfo{author}{Kuan, C.Y.}, \bibinfo{author}{Wu, H.}, \bibinfo{author}{Arora, S.}, \bibinfo{author}{Chang, K.W.}, \bibinfo{author}{Shi, J.}, \bibinfo{author}{Peng, Y.}, et~al., \bibinfo{year}{2024}.
\newblock \bibinfo{title}{{Dynamic-Superb}: Towards a dynamic, collaborative, and comprehensive instruction-tuning benchmark for speech}, in: \bibinfo{booktitle}{Proc. ICASSP}.
\newblock \DOIprefix\doi{10.1109/ICASSP48485.2024.10448257}.
\bibitem[{{Iranzo-Sánchez} et~al.(2020){Iranzo-Sánchez}, {Silvestre-Cerdà}, {Jorge}, {Roselló}, {Giménez}, {Sanchis}, {Civera} and {Juan}}]{jairsan2020a}
\bibinfo{author}{{Iranzo-Sánchez}, J.}, \bibinfo{author}{{Silvestre-Cerdà}, J.A.}, \bibinfo{author}{{Jorge}, J.}, \bibinfo{author}{{Roselló}, N.}, \bibinfo{author}{{Giménez}, A.}, \bibinfo{author}{{Sanchis}, A.}, \bibinfo{author}{{Civera}, J.}, \bibinfo{author}{{Juan}, A.}, \bibinfo{year}{2020}.
\newblock \bibinfo{title}{{Europarl-ST}: A multilingual corpus for speech translation of parliamentary debates}, in: \bibinfo{booktitle}{Proc. ICASSP}.
\newblock \DOIprefix\doi{10.1109/ICASSP40776.2020.9054626}.
\bibitem[{Le~Scao et~al.(2023)Le~Scao, Fan, Akiki, Pavlick, Ili{\'c}, Hesslow, Castagn{\'e}, Luccioni, Yvon, Gall{\'e} et~al.}]{le2023bloom}
\bibinfo{author}{Le~Scao, T.}, \bibinfo{author}{Fan, A.}, \bibinfo{author}{Akiki, C.}, \bibinfo{author}{Pavlick, E.}, \bibinfo{author}{Ili{\'c}, S.}, \bibinfo{author}{Hesslow, D.}, \bibinfo{author}{Castagn{\'e}, R.}, \bibinfo{author}{Luccioni, A.S.}, \bibinfo{author}{Yvon, F.}, \bibinfo{author}{Gall{\'e}, M.}, et~al., \bibinfo{year}{2023}.
\newblock \bibinfo{title}{Bloom: A {176B}-parameter open-access multilingual language model} \URLprefix \url{https://api.semanticscholar.org/CorpusID:253420279}.
\bibitem[{Lester et~al.(2021)Lester, Al-Rfou and Constant}]{Lester2021ThePO}
\bibinfo{author}{Lester, B.}, \bibinfo{author}{Al-Rfou, R.}, \bibinfo{author}{Constant, N.}, \bibinfo{year}{2021}.
\newblock \bibinfo{title}{The power of scale for parameter-efficient prompt tuning}, in: \bibinfo{booktitle}{Proc. EMNLP}.
\newblock \URLprefix \url{https://api.semanticscholar.org/CorpusID:233296808}.
\bibitem[{Li et~al.(2023)Li, Li, Savarese and Hoi}]{li2023blip}
\bibinfo{author}{Li, J.}, \bibinfo{author}{Li, D.}, \bibinfo{author}{Savarese, S.}, \bibinfo{author}{Hoi, S.}, \bibinfo{year}{2023}.
\newblock \bibinfo{title}{{BLIP-2}: Bootstrapping language-image pre-training with frozen image encoders and large language models}, in: \bibinfo{booktitle}{Proc. ICML}.
\newblock \URLprefix \url{https://api.semanticscholar.org/CorpusID:256390509}.
\bibitem[{Liu et~al.(2022)Liu, Ji, Fu, Tam, Du, Yang and Tang}]{liu-etal-2022-p}
\bibinfo{author}{Liu, X.}, \bibinfo{author}{Ji, K.}, \bibinfo{author}{Fu, Y.}, \bibinfo{author}{Tam, W.}, \bibinfo{author}{Du, Z.}, \bibinfo{author}{Yang, Z.}, \bibinfo{author}{Tang, J.}, \bibinfo{year}{2022}.
\newblock \bibinfo{title}{{P}-tuning: Prompt tuning can be comparable to fine-tuning across scales and tasks}, in: \bibinfo{booktitle}{Proc. {ACL} {(2)}}.
\newblock \URLprefix \url{https://aclanthology.org/2022.acl-short.8}.
\bibitem[{Liu et~al.(2023)Liu, Zheng, Du, Ding, Qian, Yang and Tang}]{liu2023gpt}
\bibinfo{author}{Liu, X.}, \bibinfo{author}{Zheng, Y.}, \bibinfo{author}{Du, Z.}, \bibinfo{author}{Ding, M.}, \bibinfo{author}{Qian, Y.}, \bibinfo{author}{Yang, Z.}, \bibinfo{author}{Tang, J.}, \bibinfo{year}{2023}.
\newblock \bibinfo{title}{{GPT} understands, too}.
\newblock \bibinfo{journal}{AI Open} \DOIprefix\doi{https://doi.org/10.1016/j.aiopen.2023.08.012}.
\bibitem[{Lugosch et~al.(2019)Lugosch, Ravanelli, Ignoto, Tomar and Bengio}]{lugosch19_interspeech}
\bibinfo{author}{Lugosch, L.}, \bibinfo{author}{Ravanelli, M.}, \bibinfo{author}{Ignoto, P.}, \bibinfo{author}{Tomar, V.S.}, \bibinfo{author}{Bengio, Y.}, \bibinfo{year}{2019}.
\newblock \bibinfo{title}{{Speech Model Pre-Training for End-to-End Spoken Language Understanding}}, in: \bibinfo{booktitle}{Proc. Interspeech}, pp. \bibinfo{pages}{814--818}.
\newblock \DOIprefix\doi{10.21437/Interspeech.2019-2396}.
\bibitem[{Ma et~al.(2024a)Ma, Liusie, Gales and Knill}]{ma2023investigating}
\bibinfo{author}{Ma, R.}, \bibinfo{author}{Liusie, A.}, \bibinfo{author}{Gales, M.J.F.}, \bibinfo{author}{Knill, K.M.}, \bibinfo{year}{2024}a.
\newblock \bibinfo{title}{Investigating the emergent audio classification ability of {ASR} foundation models}.
\newblock \href{http://arxiv.org/abs/2311.09363}{{\tt arXiv:2311.09363}}.
\bibitem[{Ma et~al.(2024b)Ma, Yang, Yang, Gao, Wang, Du, Yu, Chen, Zheng, Zhang and Chen}]{ma2024embarrassingly}
\bibinfo{author}{Ma, Z.}, \bibinfo{author}{Yang, G.}, \bibinfo{author}{Yang, Y.}, \bibinfo{author}{Gao, Z.}, \bibinfo{author}{Wang, J.}, \bibinfo{author}{Du, Z.}, \bibinfo{author}{Yu, F.}, \bibinfo{author}{Chen, Q.}, \bibinfo{author}{Zheng, S.}, \bibinfo{author}{Zhang, S.}, \bibinfo{author}{Chen, X.}, \bibinfo{year}{2024}b.
\newblock \bibinfo{title}{An embarrassingly simple approach for {LLM} with strong {ASR} capacity}.
\newblock \href{http://arxiv.org/abs/2402.08846}{{\tt arXiv:2402.08846}}.
\bibitem[{Maaz et~al.(2023)Maaz, Rasheed, Khan and Khan}]{Maaz2023VideoChatGPT}
\bibinfo{author}{Maaz, M.}, \bibinfo{author}{Rasheed, H.}, \bibinfo{author}{Khan, S.}, \bibinfo{author}{Khan, F.S.}, \bibinfo{year}{2023}.
\newblock \bibinfo{title}{{Video-ChatGPT}: Towards detailed video understanding via large vision and language models}.
\newblock \href{http://arxiv.org/abs/2306.05424}{{\tt arXiv:2306.05424}}.
\bibitem[{Muennighoff et~al.(2023)Muennighoff, Wang, Sutawika, Roberts, Biderman, Scao, Bari, Shen, Yong, Schoelkopf, Tang, Radev, Aji, Almubarak, Albanie, Alyafeai, Webson, Raff and Raffel}]{muennighoff2022crosslingual}
\bibinfo{author}{Muennighoff, N.}, \bibinfo{author}{Wang, T.}, \bibinfo{author}{Sutawika, L.}, \bibinfo{author}{Roberts, A.}, \bibinfo{author}{Biderman, S.}, \bibinfo{author}{Scao, T.L.}, \bibinfo{author}{Bari, M.S.}, \bibinfo{author}{Shen, S.}, \bibinfo{author}{Yong, Z.X.}, \bibinfo{author}{Schoelkopf, H.}, \bibinfo{author}{Tang, X.}, \bibinfo{author}{Radev, D.}, \bibinfo{author}{Aji, A.F.}, \bibinfo{author}{Almubarak, K.}, \bibinfo{author}{Albanie, S.}, \bibinfo{author}{Alyafeai, Z.}, \bibinfo{author}{Webson, A.}, \bibinfo{author}{Raff, E.}, \bibinfo{author}{Raffel, C.}, \bibinfo{year}{2023}.
\newblock \bibinfo{title}{Crosslingual generalization through multitask finetuning}.
\newblock \href{http://arxiv.org/abs/2211.01786}{{\tt arXiv:2211.01786}}.
\bibitem[{Ouyang et~al.(2022)Ouyang, Wu, Jiang, Almeida, Wainwright, Mishkin, Zhang, Agarwal, Slama, Ray et~al.}]{ouyang2022training}
\bibinfo{author}{Ouyang, L.}, \bibinfo{author}{Wu, J.}, \bibinfo{author}{Jiang, X.}, \bibinfo{author}{Almeida, D.}, \bibinfo{author}{Wainwright, C.}, \bibinfo{author}{Mishkin, P.}, \bibinfo{author}{Zhang, C.}, \bibinfo{author}{Agarwal, S.}, \bibinfo{author}{Slama, K.}, \bibinfo{author}{Ray, A.}, et~al., \bibinfo{year}{2022}.
\newblock \bibinfo{title}{Training language models to follow instructions with human feedback}.
\newblock \bibinfo{journal}{Proc. NeurIPS} \URLprefix \url{https://openreview.net/forum?id=TG8KACxEON}.
\bibitem[{Pallet et~al.(1990)Pallet, Fisher and Fiscus}]{115546}
\bibinfo{author}{Pallet, D.}, \bibinfo{author}{Fisher, W.}, \bibinfo{author}{Fiscus, J.}, \bibinfo{year}{1990}.
\newblock \bibinfo{title}{Tools for the analysis of benchmark speech recognition tests}, in: \bibinfo{booktitle}{Proc. ICASSP}.
\newblock \DOIprefix\doi{10.1109/ICASSP.1990.115546}.
\bibitem[{Panayotov et~al.(2015)Panayotov, Chen, Povey and Khudanpur}]{7178964}
\bibinfo{author}{Panayotov, V.}, \bibinfo{author}{Chen, G.}, \bibinfo{author}{Povey, D.}, \bibinfo{author}{Khudanpur, S.}, \bibinfo{year}{2015}.
\newblock \bibinfo{title}{{LibriSpeech: an {ASR} corpus based on public domain audio books}}, in: \bibinfo{booktitle}{Proc. ICASSP}.
\newblock \DOIprefix\doi{10.1109/ICASSP.2015.7178964}.
\bibitem[{Papineni et~al.(2002)Papineni, Roukos, Ward and Zhu}]{papineni2002bleu}
\bibinfo{author}{Papineni, K.}, \bibinfo{author}{Roukos, S.}, \bibinfo{author}{Ward, T.}, \bibinfo{author}{Zhu, W.J.}, \bibinfo{year}{2002}.
\newblock \bibinfo{title}{{BLEU}: a method for automatic evaluation of machine translation}, in: \bibinfo{booktitle}{Proc. {ACL}}.
\newblock \URLprefix \url{https://aclanthology.org/P02-1040/}.
\bibitem[{Radford et~al.(2021)Radford, Kim, Hallacy, Ramesh, Goh, Agarwal, Sastry, Askell, Mishkin, Clark et~al.}]{radford2021learning}
\bibinfo{author}{Radford, A.}, \bibinfo{author}{Kim, J.W.}, \bibinfo{author}{Hallacy, C.}, \bibinfo{author}{Ramesh, A.}, \bibinfo{author}{Goh, G.}, \bibinfo{author}{Agarwal, S.}, \bibinfo{author}{Sastry, G.}, \bibinfo{author}{Askell, A.}, \bibinfo{author}{Mishkin, P.}, \bibinfo{author}{Clark, J.}, et~al., \bibinfo{year}{2021}.
\newblock \bibinfo{title}{Learning transferable visual models from natural language supervision}, in: \bibinfo{booktitle}{Proc. ICML}, pp. \bibinfo{pages}{8748--8763}.
\newblock \URLprefix \url{https://api.semanticscholar.org/CorpusID:231591445}.
\bibitem[{Radford et~al.(2023)Radford, Kim, Xu, Brockman, Mcleavey and Sutskever}]{pmlr-v202-radford23a}
\bibinfo{author}{Radford, A.}, \bibinfo{author}{Kim, J.W.}, \bibinfo{author}{Xu, T.}, \bibinfo{author}{Brockman, G.}, \bibinfo{author}{Mcleavey, C.}, \bibinfo{author}{Sutskever, I.}, \bibinfo{year}{2023}.
\newblock \bibinfo{title}{Robust speech recognition via large-scale weak supervision}, in: \bibinfo{booktitle}{Proc. ICML}.
\newblock \URLprefix \url{https://proceedings.mlr.press/v202/radford23a.html}.
\bibitem[{Schick and Sch{\"u}tze(2020)}]{Schick2020ExploitingCF}
\bibinfo{author}{Schick, T.}, \bibinfo{author}{Sch{\"u}tze, H.}, \bibinfo{year}{2020}.
\newblock \bibinfo{title}{Exploiting cloze-questions for few-shot text classification and natural language inference}, in: \bibinfo{booktitle}{Proc. {EACL}}.
\newblock \URLprefix \url{https://api.semanticscholar.org/CorpusID:210838924}.
\bibitem[{Shin et~al.(2020)Shin, Razeghi, IV, Wallace and Singh}]{autoprompt:emnlp20}
\bibinfo{author}{Shin, T.}, \bibinfo{author}{Razeghi, Y.}, \bibinfo{author}{IV, R.L.L.}, \bibinfo{author}{Wallace, E.}, \bibinfo{author}{Singh, S.}, \bibinfo{year}{2020}.
\newblock \bibinfo{title}{{AutoPrompt}: Eliciting knowledge from language models with automatically generated prompts}, in: \bibinfo{booktitle}{Proc. EMNLP}.
\newblock \URLprefix \url{https://api.semanticscholar.org/CorpusID:226222232}.
\bibitem[{Tang et~al.(2024)Tang, Yu, Sun, Chen, Tan, Li, Lu, MA and Zhang}]{tang2024salmonn}
\bibinfo{author}{Tang, C.}, \bibinfo{author}{Yu, W.}, \bibinfo{author}{Sun, G.}, \bibinfo{author}{Chen, X.}, \bibinfo{author}{Tan, T.}, \bibinfo{author}{Li, W.}, \bibinfo{author}{Lu, L.}, \bibinfo{author}{MA, Z.}, \bibinfo{author}{Zhang, C.}, \bibinfo{year}{2024}.
\newblock \bibinfo{title}{{SALMONN}: Towards generic hearing abilities for large language models}, in: \bibinfo{booktitle}{Proc. ICLR}.
\newblock \URLprefix \url{https://openreview.net/forum?id=14rn7HpKVk}.
\bibitem[{Touvron et~al.(2023a)Touvron, Lavril, Izacard, Martinet, Lachaux, Lacroix, Rozière, Goyal, Hambro, Azhar, Rodriguez, Joulin, Grave and Lample}]{touvron2023llama}
\bibinfo{author}{Touvron, H.}, \bibinfo{author}{Lavril, T.}, \bibinfo{author}{Izacard, G.}, \bibinfo{author}{Martinet, X.}, \bibinfo{author}{Lachaux, M.A.}, \bibinfo{author}{Lacroix, T.}, \bibinfo{author}{Rozière, B.}, \bibinfo{author}{Goyal, N.}, \bibinfo{author}{Hambro, E.}, \bibinfo{author}{Azhar, F.}, \bibinfo{author}{Rodriguez, A.}, \bibinfo{author}{Joulin, A.}, \bibinfo{author}{Grave, E.}, \bibinfo{author}{Lample, G.}, \bibinfo{year}{2023}a.
\newblock \bibinfo{title}{{LLaMA}: Open and efficient foundation language models}.
\newblock \href{http://arxiv.org/abs/2302.13971}{{\tt arXiv:2302.13971}}.
\bibitem[{Touvron et~al.(2023b)Touvron, Martin, Stone, Albert, Almahairi, Babaei, Bashlykov, Batra, Bhargava, Bhosale et~al.}]{touvron2023llama2}
\bibinfo{author}{Touvron, H.}, \bibinfo{author}{Martin, L.}, \bibinfo{author}{Stone, K.}, \bibinfo{author}{Albert, P.}, \bibinfo{author}{Almahairi, A.}, \bibinfo{author}{Babaei, Y.}, \bibinfo{author}{Bashlykov, N.}, \bibinfo{author}{Batra, S.}, \bibinfo{author}{Bhargava, P.}, \bibinfo{author}{Bhosale, S.}, et~al., \bibinfo{year}{2023}b.
\newblock \bibinfo{title}{{LLaMa 2}: Open foundation and fine-tuned chat models}.
\newblock \href{http://arxiv.org/abs/2307.09288}{{\tt arXiv:2307.09288}}.
\bibitem[{Victor et~al.(2022)Victor, Albert, Colin, Stephen, Lintang, Zaid, Antoine, Arnaud, Arun, Manan et~al.}]{victor2022multitask}
\bibinfo{author}{Victor, S.}, \bibinfo{author}{Albert, W.}, \bibinfo{author}{Colin, R.}, \bibinfo{author}{Stephen, B.}, \bibinfo{author}{Lintang, S.}, \bibinfo{author}{Zaid, A.}, \bibinfo{author}{Antoine, C.}, \bibinfo{author}{Arnaud, S.}, \bibinfo{author}{Arun, R.}, \bibinfo{author}{Manan, D.}, et~al., \bibinfo{year}{2022}.
\newblock \bibinfo{title}{Multitask prompted training enables zero-shot task generalization}, in: \bibinfo{booktitle}{Proc. ICLR}.
\newblock \URLprefix \url{https://openreview.net/forum?id=9Vrb9D0WI4}.
\bibitem[{Wang et~al.(2023)Wang, Chen, Wu, Zhang, Zhou, Liu, Chen, Liu, Wang, Li, He, Zhao and Wei}]{wang2023neural}
\bibinfo{author}{Wang, C.}, \bibinfo{author}{Chen, S.}, \bibinfo{author}{Wu, Y.}, \bibinfo{author}{Zhang, Z.}, \bibinfo{author}{Zhou, L.}, \bibinfo{author}{Liu, S.}, \bibinfo{author}{Chen, Z.}, \bibinfo{author}{Liu, Y.}, \bibinfo{author}{Wang, H.}, \bibinfo{author}{Li, J.}, \bibinfo{author}{He, L.}, \bibinfo{author}{Zhao, S.}, \bibinfo{author}{Wei, F.}, \bibinfo{year}{2023}.
\newblock \bibinfo{title}{Neural codec language models are zero-shot text to speech synthesizers}.
\newblock \href{http://arxiv.org/abs/2301.02111}{{\tt arXiv:2301.02111}}.
\bibitem[{Watanabe et~al.(2018)Watanabe, Hori, Karita, Hayashi, Nishitoba, Unno, {Enrique Yalta Soplin}, Heymann, Wiesner, Chen, Renduchintala and Ochiai}]{Watanabe2018ESPnet}
\bibinfo{author}{Watanabe, S.}, \bibinfo{author}{Hori, T.}, \bibinfo{author}{Karita, S.}, \bibinfo{author}{Hayashi, T.}, \bibinfo{author}{Nishitoba, J.}, \bibinfo{author}{Unno, Y.}, \bibinfo{author}{{Enrique Yalta Soplin}, N.}, \bibinfo{author}{Heymann, J.}, \bibinfo{author}{Wiesner, M.}, \bibinfo{author}{Chen, N.}, \bibinfo{author}{Renduchintala, A.}, \bibinfo{author}{Ochiai, T.}, \bibinfo{year}{2018}.
\newblock \bibinfo{title}{{ESPnet}: {E}nd-to-end speech processing toolkit}, in: \bibinfo{booktitle}{Proc. Interspeech}.
\newblock \DOIprefix\doi{10.21437/Interspeech.2018-1456}.
\bibitem[{Wei et~al.(2022)Wei, Tay, Bommasani, Raffel, Zoph, Borgeaud, Yogatama, Bosma, Zhou, Metzler, Chi, Hashimoto, Vinyals, Liang, Dean and Fedus}]{DBLP:journals/tmlr/WeiTBRZBYBZMCHVLDF22}
\bibinfo{author}{Wei, J.}, \bibinfo{author}{Tay, Y.}, \bibinfo{author}{Bommasani, R.}, \bibinfo{author}{Raffel, C.}, \bibinfo{author}{Zoph, B.}, \bibinfo{author}{Borgeaud, S.}, \bibinfo{author}{Yogatama, D.}, \bibinfo{author}{Bosma, M.}, \bibinfo{author}{Zhou, D.}, \bibinfo{author}{Metzler, D.}, \bibinfo{author}{Chi, E.H.}, \bibinfo{author}{Hashimoto, T.}, \bibinfo{author}{Vinyals, O.}, \bibinfo{author}{Liang, P.}, \bibinfo{author}{Dean, J.}, \bibinfo{author}{Fedus, W.}, \bibinfo{year}{2022}.
\newblock \bibinfo{title}{Emergent abilities of large language models}.
\newblock \bibinfo{journal}{Trans. Mach. Learn. Res.} \bibinfo{volume}{2022}.
\newblock \URLprefix \url{https://openreview.net/forum?id=yzkSU5zdwD}.
\bibitem[{Wolf et~al.(2020)Wolf, Debut, Sanh, Chaumond, Delangue, Moi, Cistac, Rault, Louf, Funtowicz et~al.}]{wolf-etal-2020-transformers}
\bibinfo{author}{Wolf, T.}, \bibinfo{author}{Debut, L.}, \bibinfo{author}{Sanh, V.}, \bibinfo{author}{Chaumond, J.}, \bibinfo{author}{Delangue, C.}, \bibinfo{author}{Moi, A.}, \bibinfo{author}{Cistac, P.}, \bibinfo{author}{Rault, T.}, \bibinfo{author}{Louf, R.}, \bibinfo{author}{Funtowicz, M.}, et~al., \bibinfo{year}{2020}.
\newblock \bibinfo{title}{Transformers: State-of-the-art natural language processing}, in: \bibinfo{booktitle}{Proc. {EMNLP} (Demos)}, pp. \bibinfo{pages}{38--45}.
\newblock \DOIprefix\doi{10.18653/v1/2020.emnlp-demos.6}.
\bibitem[{Wu et~al.(2023)Wu, Gaur, Chen, Zhou, Zhu, Wang, Li, Liu, Ren, Liu and Wu}]{10389705}
\bibinfo{author}{Wu, J.}, \bibinfo{author}{Gaur, Y.}, \bibinfo{author}{Chen, Z.}, \bibinfo{author}{Zhou, L.}, \bibinfo{author}{Zhu, Y.}, \bibinfo{author}{Wang, T.}, \bibinfo{author}{Li, J.}, \bibinfo{author}{Liu, S.}, \bibinfo{author}{Ren, B.}, \bibinfo{author}{Liu, L.}, \bibinfo{author}{Wu, Y.}, \bibinfo{year}{2023}.
\newblock \bibinfo{title}{On decoder-only architecture for speech-to-text and large language model integration}, in: \bibinfo{booktitle}{Proc. ASRU}.
\newblock \DOIprefix\doi{10.1109/ASRU57964.2023.10389705}.
\bibitem[{Xu et~al.(2024)Xu, Kim, Sharaf and Awadalla}]{xu2024a}
\bibinfo{author}{Xu, H.}, \bibinfo{author}{Kim, Y.J.}, \bibinfo{author}{Sharaf, A.}, \bibinfo{author}{Awadalla, H.H.}, \bibinfo{year}{2024}.
\newblock \bibinfo{title}{A paradigm shift in machine translation: Boosting translation performance of large language models}, in: \bibinfo{booktitle}{Proc. ICLR}.
\newblock \URLprefix \url{https://openreview.net/forum?id=farT6XXntP}.
\bibitem[{Yang et~al.(2015)Yang, Yih and Meek}]{yang-etal-2015-wikiqa}
\bibinfo{author}{Yang, Y.}, \bibinfo{author}{Yih, W.t.}, \bibinfo{author}{Meek, C.}, \bibinfo{year}{2015}.
\newblock \bibinfo{title}{{W}iki{QA}: A challenge dataset for open-domain question answering}, in: \bibinfo{booktitle}{Proc. EMNLP}.
\newblock \DOIprefix\doi{10.18653/v1/D15-1237}.
\bibitem[{Yi et~al.(2021)Yi, Zhou and Xu}]{9398531}
\bibinfo{author}{Yi, C.}, \bibinfo{author}{Zhou, S.}, \bibinfo{author}{Xu, B.}, \bibinfo{year}{2021}.
\newblock \bibinfo{title}{Efficiently fusing pretrained acoustic and linguistic encoders for low-resource speech recognition}.
\newblock \bibinfo{journal}{{IEEE} Signal Process. Lett.} \bibinfo{volume}{28}, \bibinfo{pages}{788--792}.
\newblock \DOIprefix\doi{10.1109/LSP.2021.3071668}.
\bibitem[{Yu et~al.(2024)Yu, Tang, Sun, Chen, Tan, Li, Lu, Ma and Zhang}]{10445874}
\bibinfo{author}{Yu, W.}, \bibinfo{author}{Tang, C.}, \bibinfo{author}{Sun, G.}, \bibinfo{author}{Chen, X.}, \bibinfo{author}{Tan, T.}, \bibinfo{author}{Li, W.}, \bibinfo{author}{Lu, L.}, \bibinfo{author}{Ma, Z.}, \bibinfo{author}{Zhang, C.}, \bibinfo{year}{2024}.
\newblock \bibinfo{title}{Connecting speech encoder and large language model for {ASR}}, in: \bibinfo{booktitle}{Proc. ICASSP}.
\newblock \DOIprefix\doi{10.1109/ICASSP48485.2024.10445874}.
\bibitem[{Zhang et~al.(2023)Zhang, Peng, Li, Zhou and Meng}]{zhang-etal-2023-sgp}
\bibinfo{author}{Zhang, X.}, \bibinfo{author}{Peng, B.}, \bibinfo{author}{Li, K.}, \bibinfo{author}{Zhou, J.}, \bibinfo{author}{Meng, H.}, \bibinfo{year}{2023}.
\newblock \bibinfo{title}{{SGP}-{TOD}: Building task bots effortlessly via schema-guided {LLM} prompting}, in: \bibinfo{booktitle}{Proc. {EMNLP} (Findings)}.
\newblock \DOIprefix\doi{10.18653/v1/2023.findings-emnlp.891}.

\end{thebibliography}

\newpage
\appendix

\section{Dat Set Statistics}
\label{stat}
The data set training and test statistics for the corpora used in the experiments are shown in Table~\ref{corpus}. The Europarl-ST data was collected from the European Parliament debate \citep{jairsan2020a}. LibriSpeech is an audiobook reading corpus \citep{7178964}. Question-answer pairs provided by \citep{tang2024salmonn} were used to augment the dev-clean set for the SQA task, where the questions and answers were generated based on transcript text using GPT3.5. The WikiQA \citep{yang-etal-2015-wikiqa} test set with synthesised speech queries provided by \citep{tang2024salmonn} was used for the SQQA task, in which the answers generated from GPT4 were used as the reference answers.
The FSC data \cite{lugosch19_interspeech} was collected from English commands commonly used for a smart home or virtual assistant, which has 31 distinct intents.

\begin{table}[h] 
\caption{Statistics of datasets used in this paper}
\label{corpus}
\centering
\setlength{\tabcolsep}{2.0mm}
\renewcommand\arraystretch{1.05}
\begin{tabular}{ l|c|c }
\Xhline{2\arrayrulewidth}
 &\multicolumn{2}{c}{Europarl-ST} \\
\hline
ASR Data Train Set &\multicolumn{2}{c}{train}\\
\ \ -Hours&\multicolumn{2}{c}{81 hours}\\
\ \ -Samples&\multicolumn{2}{c}{34K}\\
\cline{2-3}
Few-shot ST Data Train sets&\multicolumn{1}{c|}{train-en-es-10h} &train-en-fr-10h\\
\ \ -Hours &\multicolumn{1}{c|}{10 hours}&{10 hours}\\
\ \ -Samples&\multicolumn{1}{c|}{4.2K} &{4.2K}\\
\cline{2-3}
ST Data Test sets&\multicolumn{1}{c|}{test-en-es} &test-en-fr\\
\ \ -Hours &\multicolumn{1}{c|}{2.9 hours}&{2.8 hours}\\
\ \ -Samples&\multicolumn{1}{c|}{1.3K} &{1.2K}\\
\hline
\hline
 &\multicolumn{2}{c}{LibriSpeech} \\
\hline
ASR Data Train set &\multicolumn{2}{c}{train-960}\\
\ \ -Hours&\multicolumn{2}{c}{960 hours}\\
\ \ -Samples&\multicolumn{2}{c}{281K}\\
\cline{2-3}
Test sets&\multicolumn{1}{c|}{test-clean / other} &dev-clean / other\\
\ \ -Hours &\multicolumn{1}{c|}{5.4 / 5.3 hours}&{5.4 / 5.1 hours}\\
\ \ -Samples& 2.6 / 2.9K &2.7 / 2.9K\\
\hline
\hline
&\multicolumn{2}{c}{Synthesised WikiQA}\\
\hline
    SQQA Data test set&\multicolumn{2}{c}{test}\\
\ \ -Hours&\multicolumn{2}{c}{0.5 hours}\\
\ \ -Samples&\multicolumn{2}{c}{0.6K}\\
\hline
\hline
&\multicolumn{2}{c}{Fluent Speech Commands (FSC)}\\
\hline
Few-shot SLU Data train set &\multicolumn{2}{c}{train-2h}\\
\ \ -Hours&\multicolumn{2}{c}{2 hours}\\
\ \ -Samples&\multicolumn{2}{c}{3.2K}\\
\cline{2-3}
    SLU Data test set&\multicolumn{2}{c}{test}\\
\ \ -Hours&\multicolumn{2}{c}{2.4 hours}\\
\ \ -Samples&\multicolumn{2}{c}{3.8K}\\
\Xhline{2\arrayrulewidth}
\end{tabular}
\end{table}

\section{Training and hyper-parameter details}
\label{hyper-detail}
For the Conformer encoder, the kernel size of the convolution module was set to 31. 
For models trained on the ASR data of Europarl-ST,  the attention dimension, feed-forward dimension, and attention heads of the Conformer encoder were set to 256, 2048, and 4. For models trained on LibriSpeech data, the attention dimension was set to 512 and 8 attention heads were used. 
The data was pre-processed following ESPnet \citep{Watanabe2018ESPnet} recipes. Following the ESPnet recipe, 80-dimensional logarithmic Mel filter bank (Mel-fbank) coefficients was used as the speech input feature.
Before the Mel-fbank features were fed into the encoder, convolutional layers were used to down-sample in time by a factor of 4. For models trained on the ASR data of Europarl-ST, the Mel-fbank features were extracted every 8~ms with a window size of 32~ms, the CTC ASR model was trained for 25 epochs using a learning rate $3\cdot10^{-3}$ with 25k warmup steps, and the Wav2Prompt-LLM, Encoder-LLM, and Flat-start Encoder-LLM converged after 20 epochs of training. In few-shot scenarios, the models were fine-tuned for 10 epochs. The number of trainable parameters of the Wav2Prompt-LLM, ASR-LLM Cascade, Encoder-LLM, Flat-start Encoder-LLM systems were 34.56~M, 33.77~M, 41.91~M, and 41.91~M, respectively.
For models trained on LibriSpeech data, the frame stride of the speech feature was 16~ms, following ESPnet recipes. The CTC ASR model was trained for 25 epochs using a learning rate $2\cdot10^{-3}$ with 40k warmup steps, and the Wav2Prompt-LLM and Encoder-LLM converged after 15 epochs of training. In the few-shot scenarios of FSC data, the models were fine-tuned for 20 epochs. The number of trainable parameters of the Wav2Prompt-LLM, ASR-LLM Cascade, and Encoder-LLM systems were 85.33~M,  83.74~M, and 100.01~M, respectively. 

Model training was performed on 2 NVIDIA A100 GPUs each with 80GB GPU memory. For the Europarl-ST, 40~M batch bins (as implemented by ESPnet) were used, and each epoch took about 45 minutes. For the LibriSpeech, 64~M batch bins (as implemented by ESPnet) were used, and each epoch took about 2 hours. During decoding, the beam size was set to 5. For ASR decoding, the repetition penalty as implemented by Huggingface \citep{wolf-etal-2020-transformers} was set to 1.5.

\section{SQA and SQQA evaluation}
\label{qq-eval}
Mistral-7B-Instruct-v0.2 was used to evaluate the accuracy of the model prediction in the SQA and SQQA tasks. The prompt used in this paper is listed in Table~\ref{Mistral}, which follows \cite{Maaz2023VideoChatGPT}. The accuracy was computed by counting the frequency with which the Mistral LLM outputs `yes'.

\begin{table}[t] 
\caption{Prompt used in this paper to evaluate speech QA task.}
\label{Mistral}
\centering
\setlength{\tabcolsep}{0.8mm}
\renewcommand\arraystretch{1.05}
\begin{tabular}{ l|l }
\Xhline{2\arrayrulewidth}
\hline
\multirow{13}{*}{Prompt} &{Please evaluate the following question-answer pair:}\\
\\
&Question: [question]\\
&Correct Answer: [answer]\\
&Predicted Answer: [prediction]\\
\\
&Provide your evaluation only as a yes/no and score where the score is an integer value \\
&between 0 and 5, with 5 indicating the highest meaningful match. Please generate the \\
&response in the form of a Python dictionary string with keys `pred' and `score', where \\
&value of `pred' is a string of `yes' or `no' and value of `score' is in INTEGER, not \\
&STRING. DO NOT PROVIDE ANY OTHER OUTPUT TEXT OR EXPLANATION.\\
&Only provide the Python dictionary string. For example, your response should look like\\
&this:\{`pred': `yes', `score': 4\}.\\
\Xhline{2\arrayrulewidth}
\end{tabular}
\end{table}

\section{Task-specific prompt templates}
\label{fixed}
The prefix and postfix text used in this paper as task-specific prompt templates are listed in Table~\ref{template}, which were designed based on the intended use of different LLM \citep{zhang-etal-2023-sgp, touvron2023llama2, le2023bloom} or related work \citep{xu2024a, he23_interspeech}.

\section{Limitations}

This paper is limited in the following aspects: First, this paper explores the use of off-the-shelf text-based LLMs, so the upper bound of performance is determined by the accessible text-based LLMs.
However, due to the limitations of computing resources, this paper explored various 7B LLMs. Further larger LLMs are challenging given our current computing resources. Moreover, Wav2Prompt relies on open-source LLMs and cannot use closed-source LLMs, such as GPT4.

Second, this paper follows the prompt tuning approach without updating the LLM parameters. Considering many systems have been built to compare with our proposed Wav2Prompt, fine-tuning the LLM parameters would greatly increase the resources required for training, which is challenging given our limited computing resources. Future work may explore the performance of Wav2Prompt when updating the LLM parameters.
However, the prompt tuning approach also has the advantage that only one LLM needs to be maintained for a series of tasks, which is very memory efficient in real-world deployment.

\begin{table*}[t!] 
\caption{Prefix and postfix text used in this paper}
\label{template}
\centering
\setlength{\tabcolsep}{0.4mm}
\renewcommand\arraystretch{1.09}
\begin{tabular}{ l|l }
\Xhline{2\arrayrulewidth}
\hline
ASR Train: Prefix &{""}\\
\hline
ASR Train: Postfix &{Repeat the above English text:}\\
\hline
\multirow{2}{*}{ST Test Llama: Prefix}&Translate this from English to [target language]:\\
&English: \\
\hline
ST Test Llama: Postfix&[target language]:\\
\hline
{ST Test Bloomz: Prefix}&""\\
\hline
ST Test Bloomz: Postfix&Translate the above English text into [target language]:\\
\hline
\multirow{19}{*}{SLU Test: Prefix}&We will show you some commands given by a user to a voice assistant \\
&like Siri or Olly. Please classify the intent of the command.\\
&There are 31 unique intents in total, which are divided into three slots:\\
&"action", "object", and "location". A slot takes on one of multiple values:\\
&the "action" slot can take on the values: "change language", [...];\\
&the "object" slot can take on the values: "none", "music", [...];\\
&the "location" slot can take on the values: "none", "kitchen", [...].\\
\\
&The format of intent is: "action\_object\_location". The list of all the\\
&intents are: "increase\_volume\_none", [...].\\
&You can first repeat the command and then think about the intent.\\ 
&Please give answers like: \{"Command": <your\_repeated\_command>, \\
&"Intent": <your\_intent\_prediction>\}. For example: [...]. The intent in \\
&your answer must match one of the intents given above. If you are  \\
&uncertain, choose the one that you think is the most likely.\\
&Here are the commands:\\
\\
&USER: \\
\hline
\multirow{2}{*}{SLU Test: Postfix}&Repeat the above English text and classify the intent: \\
&ASSISTANT:\\
\hline
\multirow{3}{*}{SQA Test: Prefix}&Give a precise and clear answer to the question based on the context. A\\
&good answer needs to contain the right information and be as short and\\
&clear as possible. Don't be verbose.\\
&CONTEXT:\\
\hline
\multirow{2}{*}{SQA Test: Postfix}&QUESTION: [question]\\
&ANSWER:\\
\hline
\multirow{3}{*}{SQQA Test: Prefix}&Give a precise and clear answer to the question. Don't be verbose. You\\
&can first repeat the question and then think about the answer. Please give\\
&answers like: \{"Question": <your\_repeated\_question>, "Answer": <your\\
&\_answer>\}. If you are not sure, leave the answer blank, like \{"Question":\\
& <your\_repeated\_question>, "Answer": ""\}.\\
&Here are the questions:\\
\\
&USER: \\
\hline
\multirow{2}{*}{SQQA Test: Postfix}&Repeat the above English text and answer the question: \\
&ASSISTANT:\\
\hline
ASR Test: Prefix &{""}\\
\hline
ASR Test: Postfix &{Resume the above English text:}\\
\Xhline{2\arrayrulewidth}
\end{tabular}
\end{table*}

Third, for tasks like SQQA, this paper only compared performance under a zero-shot scenario because the synthesised test set was provided by previous work of \citep{tang2024salmonn}, and we do not have a powerful speech synthesis system available. Fourth, limited by training data and computing resources, we were not able to train our Wav2Prompt as extensively as some pre-trained speech models like Whisper \citep{pmlr-v202-radford23a}, but we have conducted extensive experiments on many corpora, including LibriSpeech, the most widely used dataset used in ASR research.
This paper mainly evaluates Wav2Prompt on the English task, including ST between two European language pairs. While we believe Wav2Prompt can also be applied to other languages, including multi-lingual tasks, the performance has not been verified and is left as future work.

Finally, this paper has validated Wav2Prompt on five unseen tasks (including two ST tasks) and also ASR task. However, due to limited resources, there are still potential applications that have not been evaluated, which is left as future work.

\section{Broader impact}
Wav2Prompt allows easy integration of spoken input with text-based LLMs and provides more use-cases for off-the-shelf text-based LLMs. Wav2Prompt provides similar performance to a conventional cascade of ASR followed by a text based LLM for a zero shot performance on a range of tasks. However since it allows E2E fine-tuning, Wav2Prompt provides much improved performance in few-shot scenarios on tasks including speech translation, spoken intent classification and spoken question answering. This ability to perform very well on a range of spoken language tasks when Wav2Prompt is initially trained only on ASR training data is beneficial in many circumstances where task-specific training data is limited. This is true for many tasks, and the issue of limited spoken training data is even more severe for under-resourced languages and hence this is a significant benefit of Wav2Prompt.
Wav2Prompt does not give rise to any additional potential biases beyond the ones directly inherited from the pre-trained LLM checkpoints and the speech training data used.

\section{Assets and licenses}
The following licenses apply to the models used in this paper:
\begin{itemize}
    \item LLaMA2: \url{https://huggingface.co/meta-llama/Llama-2-7b-hf/blob/main/LICENSE.txt} applies to  LLaMA-2-7B and Vicuna-7B-1.5.
    \item Apache-2.0: \url{https://www.apache.org/licenses/LICENSE-2.0} applies to Mistral-7B-Instruct-v0.2.
    \item BigScience RAIL License v1.0: \url{https://huggingface.co/spaces/bigscience/license} applies to  BLOOMZ-7B1.
\end{itemize}

The following licenses apply to the datasets used in this paper:
\begin{itemize}
    \item CC BY-NC 4.0: \url{https://spdx.org/licenses/CC-BY-NC-4.0} applies to Europarl-ST data.
    \item CC BY 4.0: \url{https://spdx.org/licenses/CC-BY-4.0} applies to LibriSpeech data.
    \item CC-BY-NC-ND-4.0: \url{https://spdx.org/licenses/CC-BY-NC-ND-4.0} applies to Fluent Speech Commands data.
\end{itemize}

The following license applies to the code and Python package used in this paper:
\begin{itemize}
    \item Apache-2.0: applies to Huggingface Transformers (\url{https://github.com/huggingface/transformers/blob/main/LICENSE}) and ESPnet (\url{https://github.com/espnet/espnet/blob/master/LICENSE}).
\end{itemize}

\end{document}